\documentclass[aps,prl,twocolumn,superscriptaddress,showpacs]{revtex4}

\usepackage{graphics}
\usepackage[final]{epsfig}
\usepackage{color}

\newcommand{\nxc}{n_{xc}}
\newcommand{\exc}{e_{xc}}

\newcommand{\vxc}{v_{xc}}
\newcommand{\Fxc}{F_{xc}}
\newcommand{\Exc}{E_{xc}}
\newcommand{\lapln}{\nabla^2n}
\newcommand{\gradn}{\nabla n}

\newcommand{\gradnsq}{\left|\nabla n\right|^2}
\newcommand{\dexc}{\delta e_{xc}}
\newcommand{\derms}{\delta e_{rms}}

\newcommand{\GGApp}{\textrm{GGA}$^{++}$}
\newcommand{\bfr}{{\bf r}}
\newcommand{\bfrp}{{\bf r^{\prime}}}
\newcommand{\be}{\begin{equation}}
\newcommand{\ee}{\end{equation}}

\begin{document}

\preprint{Cancio et al., preprint 2005}

\title{
Beyond the local approximation to exchange and correlation: 
the role of the Laplacian of the density in the energy density of Si.
}
%
%

\author{Antonio C.~Cancio}
\email[]{accancio@bsu.edu}
\affiliation{Department of Physics and Astronomy,
Ball State University, Muncie, IN 47304}

\author{M.~Y.~Chou}
\affiliation{Department of Physics,
Georgia Institute of Technology, Atlanta GA, 30332}

\date{\today}
%
%
\begin{abstract}
We model the exchange-correlation (XC) energy density of the Si crystal and
atom as calculated by variational Monte Carlo (VMC) methods
with a gradient analysis beyond the local density approximation (LDA).
We find the Laplacian of the density to be an excellent predictor
of the discrepancy between VMC and LDA energy densities in each system.
A simple Laplacian-based correction to the LDA energy density is developed
by means of a least square fit to the VMC XC energy density for the crystal,
which fits the homogeneous electron gas and Si atom without further effort.
\end{abstract}

\pacs{71.15.Mb, 31.15.Ew, 71.10.Ca}

\maketitle
            The crucial ingredient of density functional theory~\cite{KS,JG}
            (DFT) is the the exchange-correlation (XC) energy 
            which incorporates the effects of many-body correlations on the 
            ground-state energy of an electronic system into
            its expression as a functional of the ground-state density.
            The success and widespread application of DFT
            in solid-state physics and quantum chemistry has been due
            to the
            remarkable accuracy of simple and efficient local and ``semilocal"
            models for this quantity,
            including the local density approximation (LDA)~\cite{KS},
            generalized gradient approximations
            (GGA's)~\cite{PW91,PBE,LYP},
            and various extensions of the GGA~\cite{BeckemGGA,PKZB,JTao}.
            These methods form a
            hierarchy of approximations in which this intrinsically nonlocal
            and as yet poorly understood functional of the density is mapped
            to a succession of increasingly complex local functions
            of the density, its gradient and related quantities.
            However, no systematic method for developing such corrections
            is known to exist, and the accuracy of current
            methods is not yet consistently at the level (roughly a
            milli-Rydberg)
            needed to characterize chemical reactions and other applications
            highly sensitive to the total energy.   

            A fruitful source of intuition and of mathematical
            constraints in the development of DFT's has been the
            analysis of the XC energy in terms of the XC
            hole, the change in density from the mean that occurs
            about an electron's position due to exchange and Coulomb
            correlations~\cite{JG}.  
            It provides a natural
            interpretation for the XC energy density and thus has
            aided in the construction of several DFT
            models~\cite{JG,nxc_GGA,BeckemGGA}.
            Despite the usefulness of the XC hole in DFT development,
            there have been few calculations of it for realistic
            systems.
            Recently, however, accurate variational Monte Carlo (VMC)
            calculations of the XC hole and
            the associated energy density have been performed for the
            Si crystal~\cite{Hood1PRL,Hood2PRL} and atom~\cite{Puzder} within
            a pseudopotential approximation.
            These calculations have provided a wealth of data for
            analysis~\cite{Cancio}, but a comprehensive understanding of their
            implication for DFT has to date been lacking.

            We present in this paper an analysis of the XC energy density
            associated with the XC hole in the Si crystal and atom
            in terms of a gradient analysis of the density.
            We find that the deviation of the XC energy density from
            the LDA model is markedly correlated with the local
            Laplacian of the density, a quantity that has been
            mostly neglected in developing DFT's, with the local
            gradient playing little or no role.
            We construct a minimal Laplacian-based model to quantify this
            relation with parameters fit to the crystal data.  The resulting
            fit captures most of the discrepancy between the VMC and
            LDA energy densities,
            and fits both the homogeneous electron gas (HEG) and Si atom cases
            with no further effort.

            A strong correlation between the Laplacian of the density and
            the XC energy density has previously been
            reported~\cite{NekoveeAll} for a
            model strongly inhomogeneous electron gas.
            However, the current work is the first time
            that such a picture has been found in the context of the
            complexities (covalent bonding, atomic orbitals, diamond structure)
            inherent in a real material, one that is paradigmatic for all
            covalently bonded systems.
            These results suggest the
            existence of a simple yet universal correlation between the
            XC hole and the local density Laplacian that should be a
            help in guiding future DFT models.

\begin{figure*}[t]
\begin{picture}(0,0)%
\epsfig{file=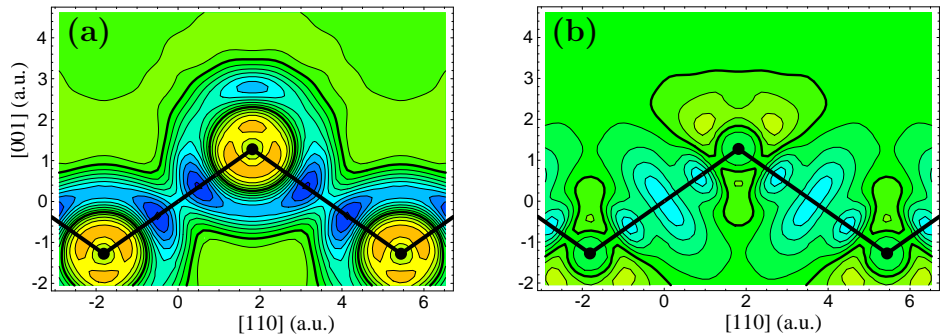}%
\end{picture}%
\setlength{\unitlength}{3947sp}%
\begingroup\makeatletter\ifx\SetFigFont\undefined%
\gdef\SetFigFont#1#2#3#4#5{%
  \reset@font\fontsize{#1}{#2pt}%
  \fontfamily{#3}\fontseries{#4}\fontshape{#5}%
  \selectfont}%
\fi\endgroup%
\begin{picture}(5830,2064)(924,-2465)
\put(1268,-623){\makebox(0,0)[lb]{\smash{{\SetFigFont{12}{14.4}{\rmdefault}{\bfdefault}{\updefault}{\color[rgb]{0,0,0}(a)}%
}}}}
\put(4315,-621){\makebox(0,0)[lb]{\smash{{\SetFigFont{12}{14.4}{\rmdefault}{\bfdefault}{\updefault}{\color[rgb]{0,0,0}(b)}%
}}}}
\end{picture}%
\caption{\label{excfig}
Comparison of DFT and VMC XC energy densities on the
(110) plane of the Si crystal.  (a) Difference in the LDA XC energy density 
and that of VMC data~\cite{Hood1PRL,Hood2PRL}.  (b) Difference between 
that of the ``\GGApp" model described in the text and VMC\@.
Contours in increments of 2$\times10^{-3}$~a.u., with thicker contour that
for zero energy difference.   Bluer regions show negative difference
and redder regions, positive.
}
\end{figure*}

\begin{figure*}
\begin{picture}(0,0)%
\epsfig{file=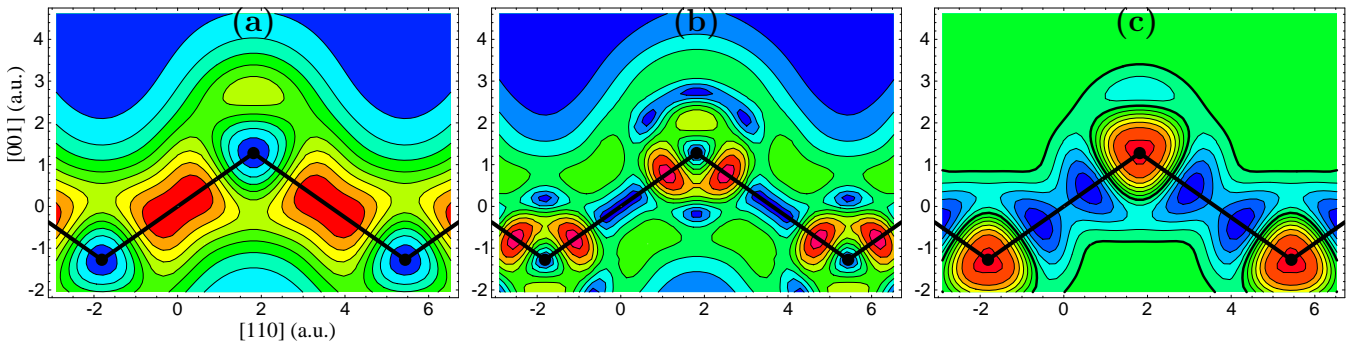}%
\end{picture}%
\setlength{\unitlength}{3947sp}%
\begingroup\makeatletter\ifx\SetFigFont\undefined%
\gdef\SetFigFont#1#2#3#4#5{%
  \reset@font\fontsize{#1}{#2pt}%
  \fontfamily{#3}\fontseries{#4}\fontshape{#5}%
  \selectfont}%
\fi\endgroup%
\begin{picture}(8399,2104)(925,-2090)
\put(7876,-136){\makebox(0,0)[lb]{\smash{{\SetFigFont{12}{14.4}{\rmdefault}{\bfdefault}{\updefault}{\color[rgb]{0,0,0}(c)}%
}}}}
\put(2326,-136){\makebox(0,0)[lb]{\smash{{\SetFigFont{12}{14.4}{\rmdefault}{\bfdefault}{\updefault}{\color[rgb]{0,0,0}(a)}%
}}}}
\put(5101,-136){\makebox(0,0)[lb]{\smash{{\SetFigFont{12}{14.4}{\rmdefault}{\bfdefault}{\updefault}{\color[rgb]{0,0,0}(b)}%
}}}}
\end{picture}%
\caption{\label{densfig}
Gradient analysis of the density of crystalline Si.  The density (a), its
gradient squared (b), and Laplacian (c) on the $(110)$ plane of the Si crystal.
Atoms and bonds outlined in black.  Shading varies from
blue (lowest) to red (highest) and contours are in increments of 0.01~a.u.~(a),
0.01~a.u.~(b), and 0.05~a.u.~(c).
In (c) the zero contour is the thicker black line.
}
\end{figure*}

      In DFT, the XC energy $\Exc$ is usually written as an integral of
      a locally defined XC energy density, $\exc$:
       \be
          \label{eqexc}
           E_{xc} = \int d^3r \; \exc(\bfr;[n]);
       \ee
      where $\exc$ is itself an unknown functional of the density $n$.
      The simplest ansatz for $\exc$ is that of the LDA
      in which the true nonlocal functional at a given point in space
      is replaced by that of the
      homogeneous electron gas (HEG) with the local value of the density:
      $\exc^{LDA}(\bfr;[n]) = \exc^{HEG}(r_s(\bfr))$, where
      $r_s = (3/4\pi n)^{1/3}$ is the Wigner-Seitz radius.  
      Corrections to the LDA are usually based on a gradient 
      expansion~\cite{Svendsen} in which the variation in the density 
      near $\bfr$, described by derivatives of $n(\bfr)$, is used to 
      modify $\exc(\bfr)$.
      GGA's add a dependence on $|\gradn(\bfr)|$, and 
      metaGGA's~\cite{PKZB}, on more complex local derivatives.
      The density Laplacian $\lapln$ occurs to the same order
      as $|\gradn|$ in the gradient expansion but is less often 
      used~\cite{BeckemGGA}.

      An intuitive picture of $\exc$ is obtained from  
      the XC hole $\nxc(\bfr,\bfrp)$, which measures the change in
      density at $\bfrp$ from the mean density $n(\bfrp)$, given the
      observation of an electron at $\bfr$.  The XC energy density 
      may be expressed in terms of the adiabatically integrated
      XC hole~\cite{Exc2nxc}:
    \be
      \label{eqadiabat}
        \exc(\bfr) = \frac{n(\bfr)}{2} \int d\lambda\;
                        \int_0^1 d^3r^{\prime}\;
                           \frac{ \nxc^{\lambda}(\bfr,\bfrp) }
                                { \left| \bfr - \bfrp  \right| }.
    \ee
(In this paper, we use hartree atomic units.)
Here, $\nxc^{\lambda}$ represents the XC hole evaluated for
a system with Coulomb coupling $\lambda e^2$ and the same ground-state
density $n(\bfr)$ as the true system.  In this formalism,
$\exc(\bfr)/n(\bfr)$
is the sum of the potential energy due to the interaction of an electron with
its own hole and the kinetic energy cost to create the hole.

    Unfortunately, $\exc$ is not uniquely definable --
    any function that integrates to zero over the system volume could be
    added to $\exc$ in Eq~[\ref{eqexc}] to generate a new ``gauge" choice
    for the energy
    density, to which the energetically relevant quantity $\Exc$
    would be invariant.
    This is implicitly done in GGA's to convert 
    any potential dependence of $\exc$ upon
    $\lapln$ to an equivalent dependence upon $\gradnsq$ alone~\cite{Svendsen}.
    On the other hand, the adiabatic method 
    is a natural, easily interpreted choice for defining $\exc$;
    moreover it is readily calculable in the VMC
    method from the expectation of the XC hole taken
    for several different values of $\lambda$~\cite{Hood1PRL}.

    To visualize the task faced
    in describing $\exc$ for a realistic system,
    we plot in Fig.~\ref{excfig}(a) the difference $\dexc$ between the
    $\exc$ of the LDA and that of the VMC calculation of
    Hood et al.~\cite{Hood1PRL,Hood2PRL} for the Si crystal in a pseudopotential
    approximation.
    The LDA predicts too deep an energy
    in the region of the Si bond, and too shallow an energy at low
    density, most obviously in the pseudo-atom core, but also,
    amplified in effect since it includes a large percentage of the unit cell
    volume, in the interstitial regions of the crystal.
    The net contribution of positive and negative errors in $\exc$
    almost exactly cancel, so that the integrated $\Exc$ in the LDA
    is essentially the same as that
    of the VMC~\cite{Hood2PRL}.
    The exact functional behavior of the energy density difference is
    quite complex.

    Figure~\ref{densfig}
    shows a gradient analysis of the Si-crystal valence electron
    density on the (110) plane.
    The gradient of the density squared $\gradnsq$, shown in
    Fig.\ref{densfig}(b), highlights the critical
    points of the density as blue regions where the gradient is nearly zero.
    It is significantly nonzero around the edges of the
    bond between two Si atoms.
    The Laplacian $\lapln$, Fig.~\ref{densfig}(c), is negative in regions
    of strong electron localization in the bond
    and positive in regions of electron depletion, such as the atom core
    and the interstitial regions.
    It has a characteristic ``butterfly shape" in the bond center, caused
    by two regions of peak density 
    located near the two Si atom valence shell peaks.

    Upon comparison of Figs.~\ref{excfig}(a) and~\ref{densfig}
    what is immediately evident is that the shape delineated
    by $\lapln$ characterizes the discrepancy between the VMC and LDA
    XC energies.  It reliably
    predicts the sign of the correction needed on a point by point basis
    throughout the unit cell,
    identifies regions of maximum error (bond and atom core),
    and reproduces key topographic features
    such as the shape of the region of maximum energy error in the bond.
    In contrast $\gradnsq$ seems
    to have little to do with the trends in energy density error.

\begin{figure}[t]
\includegraphics{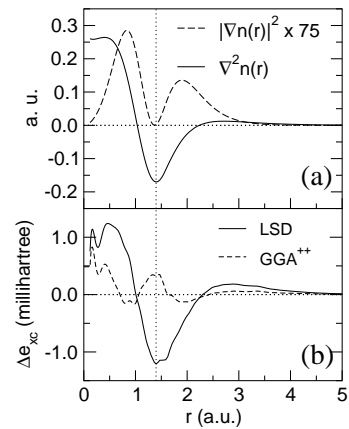}
\caption{\label{Siatomfig} Gradient analysis and \GGApp\
fit of $\exc$ for the Si pseudo-atom.  (a) Gradient squared and Laplacian of
the density as a function of radial distance from the atom core. (b) Difference
between $\exc$ of the LSD and \GGApp\ models and that obtained from
VMC~\cite{Puzder}.
}
\end{figure}

       VMC calculations of $\exc$
       have recently been performed for the valence shell of the Si atom
       in a pseudopotential model~\cite{Puzder}.
       These allow us to verify the trends demonstrated in the
       gradient analysis of the crystal in a system that lacks
       bonds, and has significantly different boundary conditions.
       Shown in Fig.~\ref{Siatomfig}(a) are $\gradnsq$ and $\lapln$
       of the Si pseudo-atom electron density versus radial distance.
       The peak of the density, indicated by the vertical dotted line,
       marks the zero of $\gradn$ and the maximum negative value of $\lapln$.
       The solid line in 3(b) shows the difference in $\exc$ between the local
       spin density (LSD)~\cite{LSD} and VMC results.
       Ignoring short-wavelength statistical fluctuations that are
       a by-product of the Monte Carlo calculation,
       a dramatic correlation of $\dexc$ with $\lapln$
       is seen, with the same qualitative trends as the crystal.

      These two examples (crystal and atom)
      demonstrate a qualitatively consistent dependence of $\exc$ upon
      the Laplacian of the density that should be quantifiable -- but not in
      the context of GGA's, which do not include a dependence on $\lapln$.
      We consider an enhanced GGA model, a ``\GGApp",
      of the form
         \be
            \label{eqggapp}
             e_{xc}^{GGA++}(r_s, s^2, l) = F_{xc}(r_s, s^2, l)\,e_{xc}^{LDA}(r_s),
         \ee
      where the correction to the LDA energy density is expressed by
      an enhancement factor $\Fxc$ dependent upon the Wigner radius $r_s$ and
      dimensionless variables
      \mbox{$l = r_s^2(\bfr) \nabla^2 n(\bfr) / n(\bfr)$} and
      \mbox{$s = r_s(\bfr) \left|\nabla n(\bfr)\right| / n(\bfr)$}.
     This \GGApp\ is fit to VMC data for the Si crystal
     by a least-squares procedure that minimizes
     the variance in the energy density from the VMC value, integrated
     over the unit cell.
    The root-mean-square error of the energy density, $\derms$, obtained in
    this way is 0.442~millihartrees for the LDA, and represents
    the average deviation from zero for the energy-difference plot
    shown in Fig.~\ref{excfig}(a).

     We have found that a form for $\Fxc$ depending only
     upon the dimensionless Laplacian $l$ provides the optimal fit to our
     data in the sense of returning the greatest degree of correction
     per fitting parameter.
     The form is given by
         \be
            \label{eqFxc}
            F_{xc}(l) = 1 + \frac{\alpha + \beta l}
                                     {1 + \gamma l}
         \ee
       with optimized fitting parameters 
       $\alpha=-0.0007$, $\beta=0.0080$, and $\gamma=0.026$.
       The fitting error $\derms$ is thereby reduced
       70\% from its LDA value to 0.132~millihartrees.
     This form potentially satisfies several known properties of the
     universal $\exc$, particularly recovering the correct value 
     in the HEG limit ($s^2=l=0$) for $\alpha=0$.  
     It behaves properly under uniform scaling
     to infinite density~\cite{Levy} but fails to include a dependence
     of $\Fxc$ on $r_s$ due to correlation.
       The smallness of the optimized value of $\alpha$ indicates that the best
       fit for the Si crystal simultaneously satisfies the HEG limit.  
       This supports the validity
       of our model as a description of a genuine physical phenomenon
       rather than a mathematical anomaly specific to Si.

        Shown in Fig.~\ref{excfig}(b) is the difference in energy
        density between our three-parameter
        \GGApp\ fit and the VMC data of Hood et al., on the same energy
        scale as the energy difference between LDA and VMC in (a),
        showing point by point what $\derms$ shows on average.
        The difference in $\exc$ has been greatly reduced everywhere
        throughout the unit cell, with the exception of the
        bond center and at the antibond point behind each bond.
       We have also tried a 5-parameter fit including terms of order
       $s^2$ and forms with
       higher order corrections, with only minimal improvement of
       $\derms$.
       In every case tried the linear coefficient $\beta$ for $l$
       remains at 0.008 to within 10\%.

      The transferability of our model can be tested by applying it to
      the Si atom data 
      of Ref.~\onlinecite{Puzder}.
      We have applied the Laplacian-only $\Fxc$ obtained from our
      fit to the crystal data without any further adjustments
      as a correction to the LSD XC energy density for the Si atom.
      This is defined similarly to Eq.~[\ref{eqggapp}],
      by $\exc^{LSD-GGA++} = \Fxc(s^2,l)\exc^{LSD}(r_s, \zeta)$, where
      $\zeta = \frac{n_\uparrow - n_\downarrow}{n}$ is the local spin
      polarization.  The result for $\dexc$ using this model is shown
      in Fig.~\ref{Siatomfig}(b);
      the overall error $\derms$ is reduced by 70\% from its LSD value,
      achieving the same reduction of error as for the crystal. 

      Our numerical results tieing $\exc$ to $\lapln$
      can be motivated qualitatively by 
      reconsidering a gradient expansion, this time for $\nxc$.
      This would use as input 
      the change in density within the length-scale of the XC hole about 
      any position, as described by local derivatives of the density,
      to correct
      the errors inherent in the LDA assumption of a locally homogeneous
      environment.
      As the Coulomb interaction is directionally invariant,
      only the change in density averaged over angle should contribute
      to this correction.
      This is precisely what is measured by $\lapln$, 
      and is unobtainable from $\gradnsq$.
      Given an $\exc$ derived from the adiabatically
      integrated XC hole, 
      one could then expect the error in the LDA model of $\exc$ 
      to be dominated by the local value of $\lapln$.

      The value of the Laplacian of the density in electronic structure
      has been noted in several other contexts.
      It has been used successfully as a 
      diagnostic tool in characterizing the electronic structure of
      molecules~\cite{Bader1}.
      Covalent bonds have been found to be
      distinguished by a negative Laplacian at the bond center, denoting
      the build-up of charge within the bond, and non-covalent ones
      by a positive $\lapln$; in addition the hour-glass pattern
      observed in the Si crystal bond is typical of other tetrahedrally bonded
      systems.
Secondly, studies of the XC potential of atoms~\cite{Engel,Umrigar} have 
pointed out
that terms in $\lapln$ are necessary to model the potential in the 
nuclear cusp and asymptotic regions.  Thus the relevance of this quantity to 
DFT extends beyond the pseudopotential models studied here to all-electron
calculations, and possibly from covalent to 
other types of chemical bonds.

    The XC potential $\vxc(\bfr)=\delta\Exc/\delta n(\bfr)$,
    necessary for a self-consistent determination of the density
    is easily obtained within the plane-wave pseudopotential formalism of
    the DFT.
    Self-consistent calculations of density and structural properties of Si
    using our \GGApp\ model show no significant deviation from the already
    reasonably good prediction of these quantities in the LDA.
    Full results will be discussed in a further paper.

      A \textit{caveat} in regard to our results is that
      our model has been fit to data obtained by a variational
      method that underestimates the correlation energy.
      The true correlation energy for each system may be lower than
      that of the VMC by about 15\%, and $\Exc$ lower by 1-2\%.
      However, within the VMC approximation, the main effect of adding
      correlation
      has been to \textit{increase} the match between the LDA error and
      $\lapln$ from
      that observed in the exchange-only case, shown in Fig.~6(a) of
      Ref.~\onlinecite{Hood2PRL}.  The effect of the addition of the
      missing correlation energy might well be to reduce further the
      discrepancy between the actual $\exc$ and a Laplacian fit.

      In summary, our fit of $\exc$ in terms of a Laplacian based
      enhancement factor $\Fxc(l)$ provides a simple model that
      has a surprisingly wide range of applicability: from the HEG to
      covalently bonded crystal to open shell atom.  This points to
      the potential
      for a Laplacian-based $\Fxc$ to make an
      excellent approximation to the true, universal one for a wide range
      of systems.
      To date, the development of GGA's and metaGGA's has emphasized
      the gradient of the density as the basic departure point for 
      the post-LDA description of DFT. 
      Our analysis indicates rather that it may be advantageous to start 
      with $\lapln$ as the key factor in going beyond the LDA.

\begin{acknowledgments}
We wish to acknowledge fruitful discussions with C. J. Umrigar and
J. P. Perdew. MYC acknowledges support from the NSF (DMR-0205328) and DOE.
\end{acknowledgments}

\bibliography{exc_fit}

\end{document}